# Comparison of capacitive and frequential readout when scaling accelerometers down from Micro- to Nano- Electro Mechanical Systems

Sébastien Hentz,[*] Laurent Duraffourg, and Eric Colinet

*CEA LETI MINATEC, 17 rue des martyrs 38054 Grenoble Cedex 9 FRANCE.*

## Abstract

This paper shows the effect of scaling silicon accelerometers down from MEMS to NEMS. It models both electronics and Brownian noise sources for both capacitive and resonant devices, and computes the minimum detectable signal attainable. Computed results are remarkably close to published experimental results. It shows the relatively low influence of the quality factor and of the beam width in the resonant case. Different scaling rules are investigated, and it appears that resonant sensing may satisfy some new application requirements, in particular for critical dimensions below a few hundreds of nm, when it is better resolved than capacitive sensing.

PACS numbers:

---

[*]Electronic address: `sebastien.hentz@cea.fr`

The question as to wether or not the general trend of miniaturization may address the need for high accuracy or low cost (and above all both, eg for mobile equipment) sensors outside of the automotive market is not clear yet[1], in spite of an extremely active research in the field. A true figure of merit should be the minimum detectable signal (MDS or resolution)[2] (Typical specifications aimed at are around 1mg resolution, with +-30g full range and 2% linearity), but most of the studies put sensitivity forward[3], as the modelling of noise sources in microsystems is in its infancy[4]. Most of the studies neglect the brownian noise[5,6]. Yet, readout techniques have improved[7,8] and smaller devices are noisier[2]. Almost none have compared the effect of scaling on the MDS for different sensing techniques. This is the goal of this paper, when scaling devices down to NEMS (making integration possible) including different noise sources, and for capacitive and frequential sensing techniques, in the context of an accelerometer.

*Capacitive detection*   The studied structure and notations are shown figure 1 . It consists in a mass M, suspended above the substrate by 4 anchors each of stiffness $k_s$, and electrostatic comb drives.

The substrate undergoing the to-be-measured acceleration $\gamma$, second Newton equation is applied to the ensemble mass+fingers. Flexure beams are considered mass less. Motion in general is assumed to be uniaxial, x is the displacement of the mass relatively to the fixed combs. Here, the electrostatic force in the comb drive due to the readout voltage has been assumed negligible. Assuming that the sensor's bandwidth is very low compared with the mass resonance pulsation $\omega_m = 2\sqrt{\frac{k_s}{M}}$, one has:

$$x(t) = \frac{\gamma(t)}{\omega_m^2} \tag{1}$$

The true sensor's sensitivity is defined by

$$S_{true} = \frac{\Delta C}{\gamma} = \frac{N\epsilon_0 e_M L_f}{\gamma}\left(\frac{1}{g_c - \frac{\gamma}{\omega_m^2}} - \frac{1}{g_c + \frac{\gamma}{\omega_m^2}}\right) \tag{2}$$

Its linearized sensitivity is then $S_{lin} = \frac{2N\epsilon_0 e_M L_f}{g_c^2 \omega_m^2}$ and the linearity specification may be chosen as $\mathcal{L} = \frac{S_{lin} - S_{true}(\gamma_{full})}{S_{lin}}$. This limit is attained at full range when the acceleration is

$$\gamma_{full} = g_c \omega_m^2 \sqrt{\frac{\mathcal{L}}{\mathcal{L}+1}} \tag{3}$$

Both thermomechanical noise and amplifier's noise will be considered throughout this paper. Temperature fluctuation, adsorption-desorption as well as defect motion noises are neglected, which is a fair approximation at our scales[9]. Using a simple dependence of the dissipation on frequency, the fluctuation-dissipation theorem and assuming that $k_B T \gg \hbar\omega$ and that the quality factor of the mass is $Q_m$, the force noise power due to thermomechanical fluctuations of the mass is[4]

$$S_f^m(\omega) = \frac{2}{\pi} k_B T \frac{M \omega_m}{Q_m} \tag{4}$$

Well below $\omega_m$, the acceleration noise power becomes

$$S_\gamma^m(\omega) = \frac{1}{M^2} S_f^m(\omega) = \frac{2}{\pi} k_B T \frac{\omega_m}{M Q_m} \tag{5}$$

For the sake of generality, the noise figure (NF) definition of the readout's amplifier will be used here[10]. Of course, the NF depends on the charge of the amplifier (eg to-be-read and parasitic capacitance) and does not represent its overall behaviour but it is accurate enough in a narrow bandwidth, and is broadly used in the RF community as a convenient quantity. The acceleration noise power brought by the amplifier is then $S_\gamma^a(\omega) = \left(10^{\frac{NF}{10}} - 1\right) S_\gamma^m(\omega)$

Assuming noise sources are uncorrelated and defining the MDS $\mathcal{R}$ as the rms acceleration noise, the total noise power is the sum of the noise powers and the displacement variance is computed as

$$\mathcal{R} = \sigma_\gamma = \sqrt{\int_0^{BW} S_\gamma^{total}(\omega) d\omega} \tag{6}$$

Replacing all variables by their value and performing the algebra gives

$$\mathcal{R} = \sqrt{\frac{2^3}{\pi}} \sqrt{k_B T} E^{\frac{1}{4}} \frac{1}{\sqrt{Q_m}} \sqrt{10^{\frac{NF}{10}} BW} \left(\frac{e_s l_s^3}{L_s^3 M^3}\right)^{\frac{1}{4}} \tag{7}$$

from where we can deduce the "resolution class" of the accelerometer $\frac{\mathcal{R}}{\gamma_{full}}$ (eq. 3). Here, the anchors' thickness $e_s$ is allowed to be different from the mass thickness, which will be commented later. As expected, the larger the mass, the higher the quality factor, the lower the measurement bandwidth and the more compliant the suspensions, the better the MDS. Of course, other considerations like feasibility and robustness limit the range of these values.

As an example, the ADXL105[11] displays $\mathcal{R} = 225\mu g$ with $BW = 1Hz$, we compute $\mathcal{R} = 261\mu g$ using this model with a typical value at this scale $NF = 10$.

*Frequential detection* The studied structure is shown figure 2. It should be seen as a generic structure, comprising a mass M amplifying its inertial force on a resonator by a factor $\Gamma$. The rotational rigidity of the anchors is neglected compared with the axial rigidity of the resonator $k_r = \frac{ES_r}{L_r}$, $S_r$ being the beam cross-section.

Considering the resonator as a Euler-Bernouilli doubly clamped beam, considering the axial force applied on the resonator $N(x,t)$ slow-varying (small sensing bandwidth), neglecting all non-linearities and noting $a(t)$ the displacement of the middle of the beam, the first mode response is computed via the Galerkin procedure[12]:

$$m_r a''(t) + b a'(t) + k_r a(t) = F(t) \tag{8}$$

with $m_r = 0.76 L_r \rho S_r$, $k_r = 379 E \frac{I_r}{L_r^3} + \frac{9.3}{L_r} N$, $\omega_N = \omega_0 \sqrt{1 + \Omega N}$, $\omega_0 = 22.4 \sqrt{\frac{E I_r}{L_r^4 S_r \rho}}$, $\Omega = 0.0246 \frac{L_r^2}{E I_r}$, $I_r = \frac{e_N l_r^3}{12}$ and $Q = \frac{m_r \omega_0}{b}$.

The resonator's absolute sensitivity is then

$$S_{true}^r = \frac{\omega_N - \omega_0}{N} = \frac{\Delta \omega}{N} = \omega_0 \frac{\sqrt{1 + \Omega N} - 1}{N} \tag{9}$$

and its linearized sensitivity $S_{lin}^r = \frac{\omega_0 \Omega}{2}$. The true sensor sensitivity is defined by

$$S_{true}(\gamma) = \frac{\Delta \omega}{\gamma} = \Gamma M S_{true}^r = \omega_0 \frac{\sqrt{1 + \Omega \Gamma M \gamma} - 1}{\gamma} \tag{10}$$

and its linearized sensitivity is $S_{lin} = \frac{\Gamma M \omega_0 \Omega}{2}$

Using the definition of the linearity specification, we have then

$$\gamma_{full} = \frac{1}{\Gamma M \Omega} \frac{4 \mathcal{L}}{(1 - \mathcal{L})^2} \tag{11}$$

.

The structure presented figure 2 is a resonating structure itself. In the absence of acceleration and damping, the system undamped eigenpulsation is $\omega_m = \sqrt{\frac{k_r d^2}{I_z}}$ where $I_z$ is the moment of inertia of the mass around the z axis.

Again, assuming a quality factor $Q_m$, the mass brownian noise power is $S_f^m(\omega) = \frac{2}{\pi} k_B T \frac{M \omega_m}{Q_m}$. The frequency noise power due to the mass fluctuations is

$$S_\omega^m(\omega) = (\Gamma S_{lin}^r)^2 S_f^m(\omega) \tag{12}$$

Following the same idea, the force noise spectral density due to thermomechanical fluctuations of the resonator is $S_f^r(\omega) = \frac{2}{\pi} k_B T \frac{m_r \omega_0}{Q}$. It may be assumed without loss of generality that the bandwidth $BW$ used by the PLL readout is very narrow compared with the $-3dB$ bandwidth of the resonator. Then following 8 the transfer function of the resonator at resonance giving the displacement versus a constant force per unit length is $H_{fx}(\omega) = \frac{Q}{k_r} = \frac{Q}{m_r \omega_0^2}$. The displacement noise power is then

$$S_x^r(\omega) = \|H_{fx}(\omega)\|^2 S_f^r(\omega) = \frac{2}{\pi} k_B T \frac{Q}{m_r \omega_0^3} \tag{13}$$

Following Robins ([13]), the frequency noise power of a closed-loop system is

$$S_\omega^r(\omega) = \left( \frac{\omega_0}{2Q} \right)^2 \frac{S_x^r(\omega_0)}{P_0} \tag{14}$$

where $P_0$ is the displacement carrier power, ie the RMS drive amplitude of the resonator $\frac{1}{2}a^2$. The latter should be driven below the hysteretic limit due to the mechanical non-linearity. Even though it is higher when operated in closed-loop[14], the open-loop stability limit

$$a_c = 1.685 \frac{l_r}{\sqrt{Q}} \qquad (15)$$

will be used here[15].

Like before, the displacement noise power brought by the amplifier is then $S_x^a(\omega) = \left(10^{\frac{NF}{10}} - 1\right) S_x^r(\omega)$ which may be expressed in frequency noise, as in 14. Under the assumption of uncorrelated sources, the total noise power is

$$S_\omega^{total}(\omega) = S_\omega^m(\omega) + S_\omega^r(\omega) + S_\omega^a(\omega) \qquad (16)$$

Computing the frequency variance as in 6, the acceleration variance is given by $\mathcal{R} = \dfrac{1}{S_{lin}} \sigma_\omega$

It appears that the mass induced noise is negligible at these scales. Neglecting it and performing the algebra gives

$$\mathcal{R} = 0.159 \sqrt{k_B T} \left(\rho E\right)^{\frac{1}{4}} \sqrt{10^{\frac{NF}{10}} BW} \frac{\sqrt{L_r} e_N}{M\Gamma} \qquad (17)$$

from where we can deduce the resolution class of the accelerometer $\dfrac{\mathcal{R}}{\gamma_{full}}$ (eq. 11).

The most surprising result here is that the MDS is independent on the quality factor, unlike an accepted idea. This is true if the error made by the frequency-tracking technique (eg PLL) is negligible and if the resonator is driven at a quality factor-dependent amplitude, like the open loop stability limit. The second surprising result is that the resolution is independent on the vibrating width of the resonator: indeed, the smaller the cross-section area, the more sensitive the resonator ($S_{lin} \propto l_r^{-2}$), but also the noisier ($S_\omega^r \propto l_r^{-2}$). What is more, the accelerometer class scales like $l_r^{-3}$. A third notable result is that the ratio full scale/resolution depends only on the resonator itself. On the other hand, large mass and amplification improve linearly the resolution (a very intuitive result) which scales like the square root of the resonator's thickness (although reducing it degrades the accelerometer class). This last remark has led to develop a process allowing different thicknesses for the mass and the resonator (process "$M\&NEMS$").

As an example, a resonant accelerometer from Berkeley University[5] displays $\mathcal{R} = 700\mu g$ with $BW = 300Hz$; we compute $\mathcal{R} = 550\mu g$ using this model with a typical value $NF = 10$, which seems very consistent without any other input than geometry.

*Comparison of frequential and capacitive detections*   The ratios of the frequential and capacitive MDS (equations 7 and 17) and resolution class are the following:

$$Ratio_{MDS} = 0.1 \rho^{\frac{1}{4}} \sqrt{Q_m} \left(\frac{e_N^2 L_r^2 L_s^3}{e_s l_s^3}\right)^{\frac{1}{4}} \frac{1}{M^{\frac{1}{4}} \Gamma} \qquad (18)$$

$$\mathcal{R}atio_c = 0.12\rho^{\frac{1}{4}}\sqrt{Q_m}\frac{(\mathcal{L}-1)^2}{\mathcal{L}(\mathcal{L}+1)}\left(\frac{e_s^3 l_s^9 L_r^{10}}{L_s^9 e_N^2 l_r^{12}}\right)^{\frac{1}{4}}\frac{g_c}{M^{\frac{1}{4}}} \qquad (19)$$

Figure 3 shows how these quantities scale with a critical dimension for different sets of technologically and physically reasonable parameters. The smaller, the more advantageous for frequential sensing, becoming better resolved than capacitive sensing for critical dimensions below a few hundreds of nm, ie even feasible with DUV lithography. On the other hand, the resolution class is quasi always better for the capacitive pickoff (up to several orders of magnitude), but its monotony depends on the scaling rules used. The choice of the pickoff will then depend on the application aimed at. There are of course many other considerations when designing a sensor, but this study tends to show that scaling devices down may be an answer to reach unprecedented resolutions, as well as low cost devices. This graph should then be used as an extremely useful design tool: it shows in particular the ranges of performances attainable with existing technologies like nanowires, e-beam or DUV lithographies. This should be experimentally validated in the near future, as devices will be fabricated with these three technologies, and in particular with thicknesses different for the mass and the resonator.

The authors acknowledge financial support from the French ANR-M&NEMS contract.

*List of captions*

Figure 1 : Capacitive accelerometer, studied structure

Figure 2 : Resonant accelerometer, studied structure

Figure 3 : Frequential/capacitive MDS and class ratios vs $l_r$, with $L_M = 150\mu$, $l_M = 120\mu$, $\Gamma = 37.5$, $L_r = 100l_r$, $e_N = 2l_r$, $g_c = l_s$, $Q_m = 10000$, $BW = 100Hz$, $\mathcal{L} = 2\%$, $d = 2\mu$, $NF = 10$. Frequency, MDS and $\gamma_{full}$ are given for the resonant cases 2,3,4

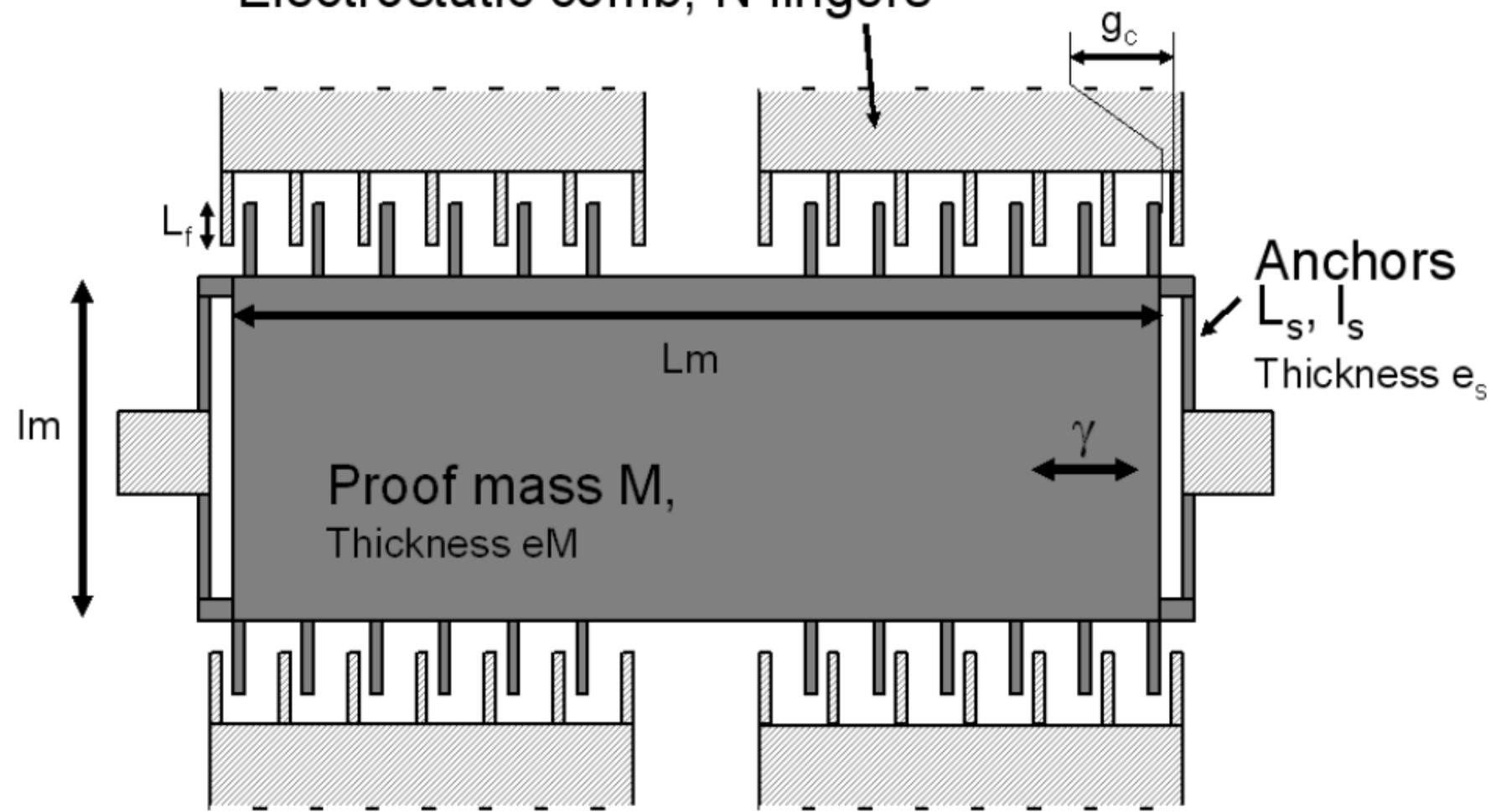

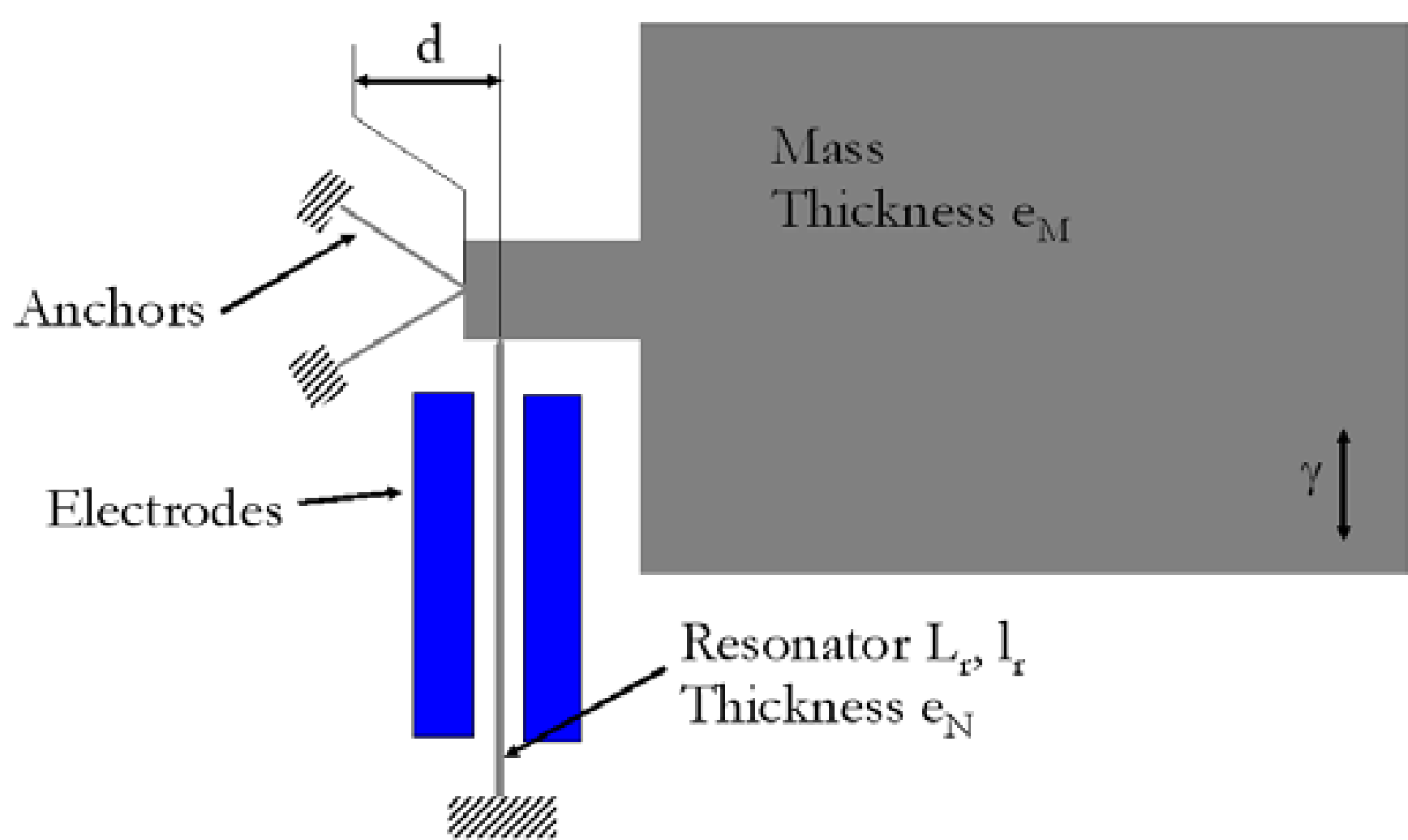

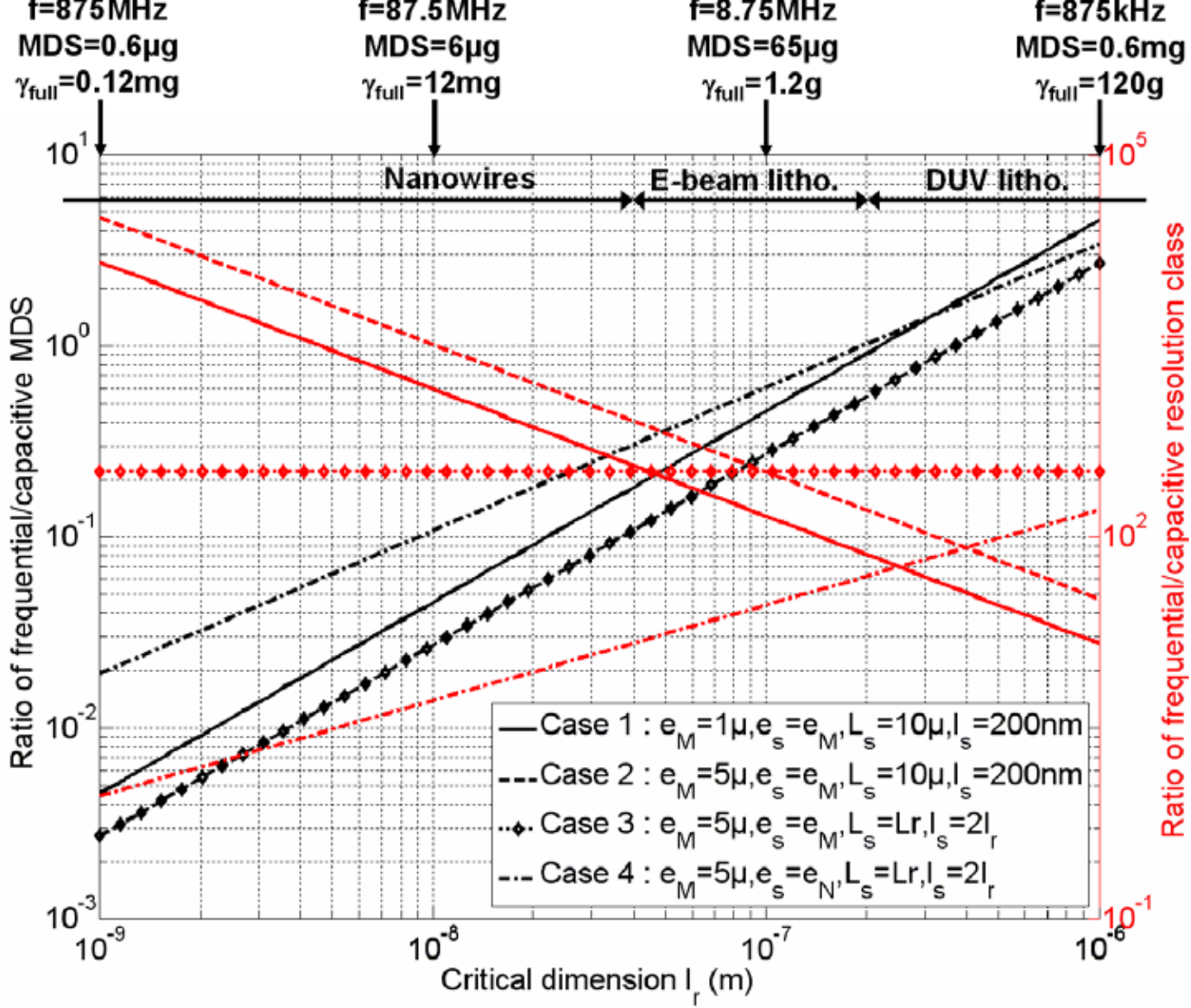